# Smeared spin-flop transition in random antiferromagnetic Ising chain

P.N. Timonin

Southern Federal University, 344090, Rostov-on-Don, Russia

*At $T = 0$ and large enough field the nearest-neighbor antiferromagnetic Ising chain undergoes first-order spin-flop transition into ferromagnetic phase. We consider its smearing under the random-bond disorder such that all independent random bonds are antiferromagnetic (AF). It is shown that ground state thermodynamics of such random AF chain can be described exactly for arbitrary distribution of AF bonds $P(J)$. Moreover, the site magnetizations of finite chains can be found analytically in this model. We consider continuous $P(J)$ which is zero above some $-J_1$ and behaves near it as $(-J_1-J)^\lambda$, $\lambda > -1$. In this case ferromagnetic phase emerges continuously in a field $H > H_c = 2J_1$. At $0 > \lambda > -1$ it has usual second-order anomalies near $H_c$ with critical indices obeying the scaling relation and depending on $\lambda$. At $\lambda > 0$ the higher-order transitions appear (third, fourth etc.) marked by the divergence of corresponding nonlinear susceptibilities. In the chains with even number of spins the intermediate "bow-tie" phase with linearly modulated AF order exists between AF and ferromagnetic ones at $J_1 < H < H_c$. Its origin can be traced to the infinite correlation length of the degenerate AF phase from which it emerges. This implies the existence of similar inhomogeneous phases with size- and form-dependent order in a number of other systems with infinite correlation length. The possibility to observe the signs of "bow-tie" phase in low-T neutron diffraction experiments is discussed.*

The influence of the frozen disorder on a first-order phase transition was first described phenomenologically by Imry and Wortis [1] who have shown that "random-temperature" disorder can diminish or even eliminate the jumps of order parameter and other variables at the transition point. Further studies have revealed the relation of such smeared transitions to that of the random field Ising model [2], established possibility for them to become a second-order ones [3-5] and to transform into a phase coexistence region instead of a sharp transition [6].

Yet our understanding of these smearing phenomena is far from exhaustive. We still have no rigorous criteria to decide which of the known outcomes of the smearing will be realized – softened jumps, phase coexistence region or a second-order transition – having only qualitative considerations on this point [7]. For the resulting second-order transition it is not known definitely are the critical indices universal or depend on the disorder parameters [3 - 5]. There is also the unexplored possibility that phase coexistence region, lacking first- and second-order transition anomalies, contains at some point the higher-order ones proper to the higher-order transitions. And we know nothing about the influence of disorder on the athermal and ground-state transitions such as spin-flop transitions in antiferromagnets (AF). At these first-order transitions some spin sublattice upturns to become (partially) parallel to external magnetic field at some critical field strength [8]. The AF Ising spin chain provides the simplest example of such transition from AF phase to ferromagnetic one at $T = 0$ and field $H = 2J$, $J$ being the absolute value of AF exchange. In the presence of random $J$ variations some segments of the chain would become ferromagnetic (F) at lower or higher $H$ values so one can expect some smearing of this first-order transition.

Fortunately, at $T = 0$ the magnetic properties of AF random-bond chain can be described exactly for arbitrary distribution of exchanges and even for arbitrary chain length. So we have unique possibility for analytical study of smeared spin-flop transition in finite samples. Here we present the results for a continuous distribution of exchanges which makes the first-order jumps completely smeared in F phase.

## 1. Distribution functions for effective random fields.

We consider short-range random-bond Ising chain with the Hamiltonian

$$\mathcal{H} = -\sum_{n=0}^{N-1} J_n S_n S_{n+1} - H \sum_{n=0}^{N-1} S_n \qquad (1)$$

Here $J_n$ are random nearest-neighbor exchanges with the identical distribution function $P(J)$, $P(J) = 0$ at $J \geq 0$.

To date there are numerous studies of random bond Ising chains either at $T=0$ or at finite $T$ [9-17]. Their common feature is the existence of definite thermodynamic limit for the main function describing their properties, that is, for the distribution of effective random fields $W_n(F)$,

$$W_n(F) = \langle \delta(F - F_n) \rangle_J ,$$

$$F_n \equiv \frac{1}{2} \ln \left( \frac{Z_n(+1)}{Z_n(-1)} \right)$$

Here $Z_n(S)$ is the partial partition function of the chain of the length $n$ (assuming the unit spacing between spins), summed over all spins except the end one $S$, angular brackets denote the average over the bond distribution function $P(J)$. The recursion relations for $Z_n(S)$,

$$Z_{n+1}(S) = \sum_{S'=\pm 1} \exp \beta S'(JS + H) Z_n(S'), \quad \beta \equiv T^{-1}$$

generate the corresponding relations for $F_n$

$$F_{n+1} = T \tanh^{-1} \left[ \tanh \beta J \tanh \beta (F_n + H) \right] \equiv U(F_n, J) \qquad (2)$$

and for $W_n(F)$

$$W_{n+1}(F) = \int \langle \delta(F - U(F', J)) \rangle_J W_n(F') dF' \qquad (3)$$

With the initial conditions $F_0 = 0$ and, correspondingly, $W_0(F) = \delta(F)$ to these equations, we can find all $W_n(F)$ and all average thermodynamic variables of our random chain. Thus, for the average magnetization of the site situated at the distance *n* from one end of a chain with *N* spins and at the distance $n' = N - n - 1$ from the other end we have

$$m_{n,N} = \left\langle \frac{\sum_{S=\pm 1} Z_n(S) S e^{\beta HS} Z_{n'}(S)}{\sum_{S=\pm 1} Z_n(S) Z_{n'}(S)} \right\rangle_J = \iint W_n(F) W_{n'}(F') \tanh \beta (F + F' + H) dF dF' \quad (4)$$

Usually, it is tedious task to find all $W_n(F)$ even at *T* = 0, so most previous studies rely heavily on the existence of the thermodynamic limit $W_\infty(F)$. In the random AF model with *P(J)*=0 at $J \geq 0$ the ground-state $W_n(F)$ can be easily found for every *n* as we show below. This does not only allow to study analytically the finite-size effects but also makes feasible the description of ground-state properties of the chains with the even number of sites, *N*. The "even chains" preserve the two-fold degeneracy of ground-state in small enough *H* which results in the infinite correlation length at *T*=0. Therefore, the boundary effects may spread throughout the whole even chain which necessitates the consideration of finite samples. Here we should note that formally one can obtain the thermodynamic limit $W_\infty(F)$ for odd and even *n* separately but they always give sensible results for the interior of odd chains only, when correlation length is finite at all *H*.

At *T* = 0 it is more convenient to consider the recursion relations for integrated probability distributions

$$C_n(F) = \int_{-\infty}^{F} W_n(F') dF'$$

Apparently,

$$C_n(-\infty) = 0, \quad C_n(\infty) = 1. \quad (5)$$

Integrating (3) we get

$$C_{n+1}(F) = \int \left\langle \vartheta[F - U(F', J)] \right\rangle_J \partial_{F'} C_n(F') dF' =$$
$$= \left\langle \vartheta[F - U(\infty, J)] \right\rangle_J + \int \left\langle \delta[F - U(F', J)] \partial_{F'} U(F', J) \right\rangle_J C_n(F') dF' \quad (6)$$

Here $\vartheta(F)$ is Heaviside's step function, $\partial_{F'} \equiv \dfrac{\partial}{\partial F'}$ and $U(\infty, J) = J$, cf. (2). As $|U(F', J)| \leq |J|$ the average in (6) is confined to the region $J^2 \leq F^2$. In this region the equation $F = U(F', J)$ has unique solution

$$F' = -H + \tanh^{-1}\left(\frac{\tanh \beta F}{\tanh \beta J}\right) \equiv V(F, J) \qquad (7)$$

Hence, we can represent delta-function in (6) as

$$\delta[F - U(F', J)] = \vartheta(F^2 - J^2) |\partial_{F'} U(F', J)|^{-1} \delta[V(F, J) - F'] \qquad (8)$$

As $sign[\partial_{F'} U(F', J)] = signJ$ we get from Eqs. (6), (8)

$$C_{n+1}(F) = Q(F) + \int \langle signJ \vartheta(F^2 - J^2) C_n[V(F, J)] \rangle_J \qquad (9)$$

$$Q(F) \equiv \langle \vartheta(F - J) \rangle_J = \int_{-\infty}^{F} P(J) dJ \qquad (10)$$

At $T = 0$ Eq. (9) greatly simplifies as here we can put $V(F, J) = -H + F signJ$, cf. (7). So Eq. (9) becomes at $T = 0$

$$C_{n+1}(F) = Q(F) + [1 - Q(|F|)] C_n(F - H) + Q(-|F|) C_n(-F - H) \qquad (11)$$

Thus we have the functional equations for the ground-state $C_n$ which in many cases can be easily solved for $n \to \infty$ at least. In particular, for discrete bond distributions when $P(J)$ is a sum of delta-functions, $Q$ is stepwise constant and the same holds for $C_n$. Then Eqs. (11) become a series of algebraic equations for the jumps of $C_n$ at some points $F_i$ the position of which is dictated by the form of Eq.(11) and by the initial condition $C_0(F) = \vartheta(F)$. In this way one may easily reproduce many known results for the ground state of various random chains obtained by other methods [9-17]. Yet here we deal with even simpler model which has not enjoyed attention previously.

## 2. Ground-state field distributions for random antiferromagnetic chains.

If $P(J) = 0$ at $J \geq 0$ then $Q(|F|) = 1$ and Eq.(11) becomes

$$C_{n+1}(F) = Q(F) + Q(-|F|)C_n(-F-H) \qquad (12)$$

Changing the variables in it, $F \to -F-H$, we get another equation

$$C_{n+1}(-F-H) = Q(-F-H) + Q(-|F+H|)C_n(F) \qquad (13)$$

Thus we have two equations for two functions, $C_n(F)$ and $\tilde{C}_n(F) = C_n(-F-H)$. In a matrix form they read

$$\mathbf{C}_{n+1} = \hat{R}\mathbf{C}_n + \mathbf{Q} \qquad (14)$$

Here

$$\mathbf{C}_n = \begin{pmatrix} C_n(F) \\ \tilde{C}_n(F) \end{pmatrix}, \mathbf{Q}_n = \begin{pmatrix} Q_n(F) \\ \tilde{Q}_n(F) \end{pmatrix}, \hat{R} = \begin{pmatrix} 0 & -R(F) \\ -\tilde{R}(F) & 0 \end{pmatrix}, R(F) \equiv Q(-|F|) \qquad (15)$$

and tilde designates the variable change, $F \to -F-H$. Note that this operation transforms a couple of functions $A(F)$ and $\tilde{A}(F)$ one into another and this is the reason for the exact solvability of Eq. (12). Also using the notations $\tilde{A}(F)$ for this variable change we may drop functions' arguments in (14) as they are the same (*F*) for all functions.

Initial conditions for Eq. (14) are

$$\mathbf{C}_0 = \begin{pmatrix} \vartheta(F) \\ \vartheta(-F-H) \end{pmatrix} \qquad (16)$$

and their solution for $n \geq 1$ is

$$\mathbf{C}_n = \hat{R}^n \mathbf{C}_0 + \sum_{k=0}^{n-1} \hat{R}^k \mathbf{Q} \qquad (17)$$

Eigenvalues of $\hat{R}$ are

$$r_\pm = \pm\sqrt{R\tilde{R}} \equiv \pm\rho \qquad (18)$$

so in the regions of *F* where $\rho < 1$ Eq. (17) can be represented as

$$\mathbf{C}_n = \mathbf{C}_\infty + \hat{R}^n(\mathbf{C}_0 - \mathbf{C}_\infty), \quad \mathbf{C}_\infty \equiv (\hat{I} - \hat{R})^{-1}\mathbf{Q} = (1-\rho^2)^{-1}(\hat{I}+\hat{R})\mathbf{Q} \qquad (19)$$

It is easy to verify that

$$\hat{R}^n = v_n^+ \rho^n \hat{I} + v_n^- \rho^{n-1}\hat{R}, \quad v_n^\pm = \frac{1}{2}\left[1 \pm (-1)^n\right] \qquad (20)$$

Thus in these regions $\mathbf{C}_n$ has definite thermodynamic limit, $\mathbf{C}_\infty$.

Yet in some regions of $F$ $\rho$ can be equal to 1 and here we have

$$\hat{R} = \begin{pmatrix} 0 & -1 \\ -1 & 0 \end{pmatrix} \equiv -\hat{\sigma}_x,$$

$$\mathbf{C}_n = \left(v_n^+ \hat{I} - v_n^- \hat{\sigma}_x\right)\mathbf{C}_0 + \frac{1}{2}\left[n(\hat{I}-\hat{\sigma}_x) + v_n^-(I+\hat{\sigma}_x)\right]\mathbf{Q} \qquad (21)$$

Further we consider the model with smooth $P(J)$ such that $P(J) = 0$ for $J > -J_1$. The behavior of $Q$, $\tilde{Q}$, $R$ and $\tilde{R}$ in this case is shown schematically in Fig. 1 for three ranges of $H$. When $H < 2J_1$ there is region $-J_1 < F < J_1 - H$ in which $\rho = 1$, while at $2J_1 < H$ $\rho < 1$ for all $F$. This and the form of $\mathbf{C}_0$ (16) predetermine the differences of $\mathbf{C}_n$ in three regions of $H$ values.

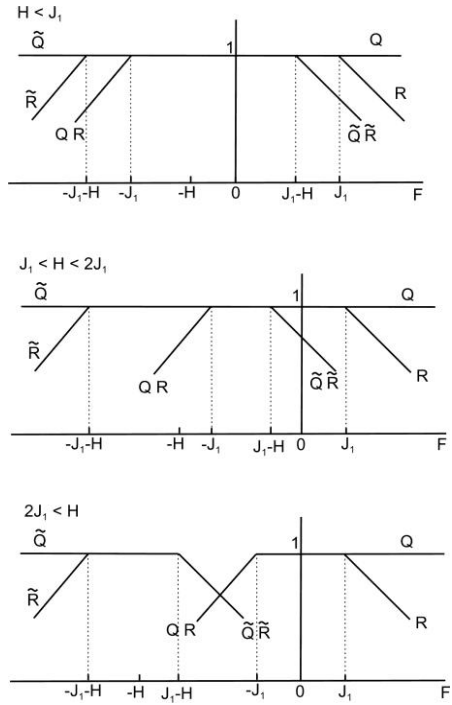

Fig.1. Schematic behavior of $Q$, $\tilde{Q}$, $R$ and $\tilde{R}$ for random AF with $P(J) = 0$ at $J > -J_1$.

From (16), (19) - (21) we get

$H < J_1$,

$$C_n(F) = v_n^- \vartheta(F+H) + v_n^+ \vartheta(F), W_n(F) = v_n^- \delta(F+H) + v_n^+ \delta(F) \quad (22)$$

$J_1 < H < 2J_1$,

$$C_n(F) = v_n^- \vartheta(F+H)\left[\rho^{n-1}(F)R(F)\vartheta(-J_1-F) + \vartheta(J_1+F)\right] + \\ + v_n^+ \vartheta(F+H-J_1)\left[1 - \vartheta(-F)\rho^n(F)\right] \quad (23)$$

$$W_n(F) = v_n^- \vartheta(F+H)\partial_F\left[\rho^{n-1}(F)R(F)\vartheta(-J_1-F)\right] - \\ - v_n^+ \vartheta(F+H-J_1)\partial_F\left[\vartheta(-F)\rho^n(F)\right] \quad (24)$$

$2J_1 < H$,

$$C_n(F) = v_n^- \left\{\vartheta(F+H)\rho^{n+1}(F) + C_\infty(F)\left[1 - \rho^{n+1}(F)\right]\right\} + \\ + v_n^+ C_\infty(F)\left[1 - \vartheta(-F)\rho^n(F)\right] \quad (25)$$

$$W_n(F) = v_n^- \left\{\partial_F\left[\vartheta(F+H)\rho^{n+1}(F)\right] + \partial_F\left[C_\infty(F)(1-\rho^{n+1}(F))\right]\right\} + \\ + v_n^+ \left\{W_\infty(F)\left[1-\rho^n(F)\right] - C_\infty(F)\partial_F\left[\vartheta(-F)\rho^n(F)\right]\right\} \quad (26)$$

In (25), (26) $C_\infty(F)$ and $W_\infty(F) = \partial_F C_\infty(F)$ are the values of $C_n(F)$ and $W_n(F)$ in the thermodynamic limit, $n \to \infty$,

$$C_\infty(F) = \vartheta(F+J_1) + \vartheta(-F-J_1)\frac{Q(1-\tilde{Q})}{1-Q\tilde{Q}} \quad (27)$$

$$W_\infty(F) = \frac{P(1-\tilde{Q}) + \tilde{P}Q(1-Q)}{(1-Q\tilde{Q})^2}\vartheta(F+J_1-H)\vartheta(-J_1-F) \quad (28)$$

Here $P = P(F)$ and $\tilde{P} = P(-F-H)$. $C_\infty(F)$ and $W_\infty(F)$ at $2J_1 < H$ for the bond distribution

$$P(J) = \vartheta(J+J_0)\vartheta(-J_1-J)(\lambda+1)\frac{(-J_1-J)^{\lambda}}{(J_0-J_1)^{\lambda+1}} \quad (29)$$

with $J_0 = 10J_1$ and $\lambda > -1$ are shown in Fig. 2.

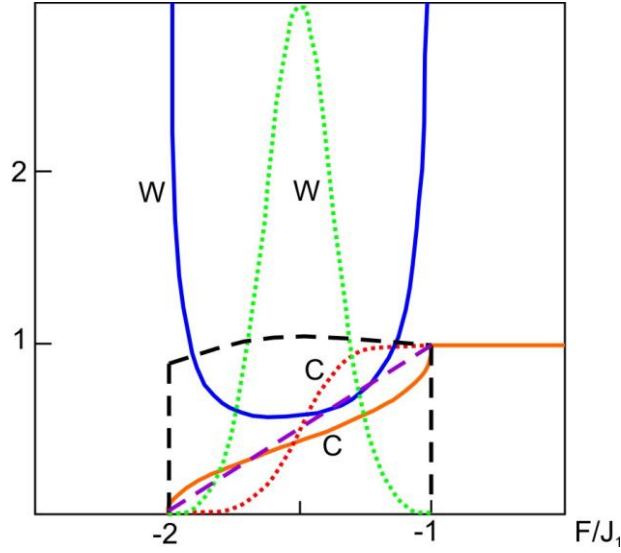

Fig. 2. $C_\infty(F)$ and $W_\infty(F)$ at $2J_1 < H$ for the bond distribution (29) with $\lambda = 2$ ( dotted lines), $\lambda = 0$ (dashed lines) and $\lambda = -0.5$ ( solid lines), $J_0$ = 10, $H$ = $3J_1$.

Eqs. (22)-(28) suffice to give full description of magnetic properties of finite random AF chains at $T$ = 0.

### 3. Magnetizations and phase transitions at *T* = 0: odd *N*.

The average ground-state magnetizations can be obtained using $T$ = 0 limit of Eq. (4),

$$m_{n,N} = \iint W_n(F)W_{n'}(F')sign(F+F'+H)dFdF', \quad sign(0)=0 \quad (30)$$

This relation can be represented in a more convenient form

$$m_{n,N} = 1 - 2\int \tilde{C}_n W_{n'} dF \quad (31)$$

In (31) the expressions (22) – (28) should be used with $\vartheta$-functions defined at zero as $\vartheta(0) = \frac{1}{2}$ to conform with $sign(0) = 0$ in (30) and the relation $sign(x) = \vartheta(x) - \vartheta(-x)$ used in derivation of (31).

The magnetization for odd and even *N* can differ drastically. The formal reason for this is that the parity of the distances of a given site from the ends of a chain ($n$ and $n' = N - n - 1$) are the same in the former case and different in the latter one.

For odd *N* we have from (22) – (31)

$$H < 2J_1, \qquad m_{n,N} = (-1)^n$$

$$H > 2J_1,$$

$$m_{n,N} = m_\infty + \int W_\infty \left\{ \left(\rho^n + \rho^{n'}\right)\left[\tilde{C}_\infty(1+\rho) - \rho\right] + \rho^{N-1}\left[\rho^2 - \tilde{C}_\infty(1+\rho^2)\right] \right\} dF$$
$$+ (-1)^n \int W_\infty \left\{ \left(\rho^n + \rho^{n'}\right)\left[\tilde{C}_\infty(1-\rho) + \rho\right] + \rho^{N-1}\left[\tilde{C}_\infty(\rho^2 - 1) - \rho^2\right] \right\} dF \qquad (32)$$

$$m_\infty = 1 - 2\int W_\infty \tilde{C}_\infty dF \qquad (33)$$

Integration in (32), (33) is limited to the interval $J_1 - H < F < -J_1$ in which $W_\infty \neq 0$, cf. (28). In this interval $\rho = \sqrt{Q\tilde{Q}} < 1$ so in the thermodynamic limit ($n, n', N \to \infty$) $m_{n,N} = m_\infty$ and we have phase transition at $H = H_c \equiv 2J_1$ from AF phase into F one. From the F-side this transition is continuous as $m_\infty$ tends to zero when $H \to H_c + 0$. Indeed, if *P(J)* vanishes or stays finite at –*J₁* then $W_\infty \to \delta(F + H/2)$, see Fig. 2, and we have

$$m_\infty \approx 1 - 2C_\infty(-H/2) = \frac{1 - Q(-H/2)}{1 + Q(-H/2)} \approx \frac{1}{2}\int_{-H/2}^{-J_1} P(J)\, dJ.$$

When *P(J)* diverges at –*J₁* we can represent $m_\infty$ as

$$m_\infty = 2 \int_{J_1-H}^{-J_1} \frac{Q - \tilde{Q}}{(1 - Q\tilde{Q})^3}\left(1 - \tilde{Q}\right)^2 P\, dF \approx 2 \int_{J_1-H}^{-J_1} P\, dF$$

With the power-law dependence of *P(J)* near –*J₁* as in Eq. (29) we have in both cases

$$m_\infty \sim (H - H_c)^{\lambda+1}, \qquad \chi \equiv \frac{\partial m_\infty}{\partial H} \sim (H - H_c)^\lambda \qquad (34)$$

So at $-1 < \lambda < 0$ there is usual second-order transition with critical indices

$$\beta = \lambda + 1, \quad \gamma = -\lambda$$

From the relation $\chi = -\dfrac{\partial^2 E}{\partial H^2}$ (E is the average energy) it follows that index $\alpha$ is equal to $\gamma$ so the usual scaling relation holds

$$\alpha + 2\beta + \gamma = 2$$

We can formally obtain indices $\nu$ and $\eta$

$$\nu = (2-\alpha)/d = 2 + \lambda$$

$$\eta = 2 - \frac{\gamma}{\nu} = \frac{3\lambda + 4}{\lambda + 2}$$

These scaling relations imply the following form of the average correlation function near $H_c$

$$G_r \equiv \langle S_n S_{n+r} \rangle_{0,J} - \langle \langle S_n \rangle_0 \langle S_{n+r} \rangle_0 \rangle_J = \frac{g(r/\xi)}{r^{2\beta/\nu}}. \qquad (35)$$

Here $\langle ... \rangle_0$ designates the average over (J-dependent) ground state(s), $\xi \sim (H - H_c)^{-\nu}$ is correlation length, $g(x)$ falls faster than any power of $x$ at $x \to \infty$. We cannot verify this prediction as the calculation of $G_r$ in F- phase is a separate task lying outside the scope of this paper.

At $\lambda > 0$ the higher-order field-derivatives of $m_\infty$ diverge so we can interpret the behavior in (34) as higher-order phase transitions (third-order for $0 < \lambda < 1$, fourth-order for $1 < \lambda < 2$, etc.). When $P(J)$ goes to zero near $-J_1$ faster than any power of $(-J_1 - J)$ we would have the infinite-order phase transition. Yet from the AF phase side there is always sharp drop of AF order parameter from 1 to zero, i.e. the first-order transition anomaly. Note also that for $\lambda = 0$ we have only a jump of linear susceptibility from zero to a finite value at $H = H_c$ as in an ordinary first-order transition. For integer $\lambda = 1, 2, ...$ only corresponding nonlinear susceptibilities experience similar jumps which have no analog among the known types of transitions.

In finite samples there is no sharp transition into F phase; instead at $H > H_c$ this phase starts to form gradually in the middle of a chain. When $P(J) \to 0$ at $J \to -J_1$ we again use $W_\infty \approx \delta(F + H/2)$ near $H_c$ to obtain

$$m_{n,N} = m_\infty \left(1 - e^{-\kappa N}\right) + (-1)^n \left[(1 - m_\infty)\left(e^{-\kappa n} + e^{-\kappa n'}\right) - e^{-\kappa N}\right]$$

$$\kappa \equiv -\ln Q(-H/2) \approx 2m_\infty \qquad (36)$$

Hence, the intervals of order $\kappa^{-1}$ length at both ends are still occupied by the (exponentially modulated) AF phase. So both phases coexist in finite chain when $N\kappa \geq 2$ $(m_\infty \geq N^{-1})$, the fraction of such AF phase being $2/N\kappa \approx (Nm_\infty)^{-1}$, while when $m_\infty < N^{-1}$ the whole chain is still in (slightly modulated) AF phase. Note also that in general $\kappa^{-1} \sim (H-H_c)^{-\beta}$ behaves differently from the correlation length $\xi \sim (H-H_c)^{-\nu}$. Similar results hold when $P(J)$ diverges at $J \to -J_1$, see Fig. 3.

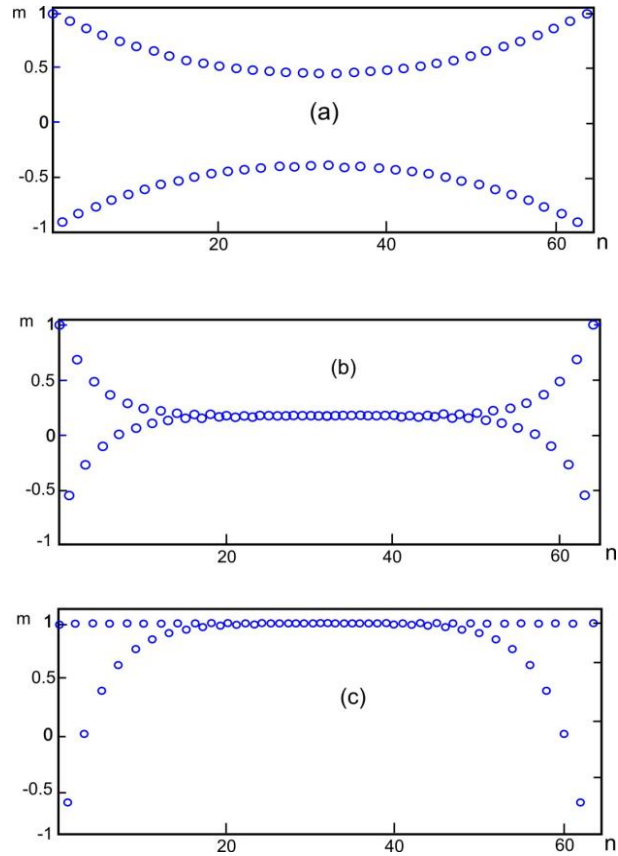

Fig. 3. Average ground-state magnetizations of odd chains with $N = 65$ and $P(J)$ from Eq.(29) with $J_0 = 10J_1$. (a) $H = 3J_1$, λ = 0.1, (b) $H = 3J_1$, λ = -0.5, (c) $H = 6J_1$, λ = -0.5.

### 3. Magnetization and phase transitions at $T = 0$: even $N$.

For even $N$ Eqs. (22) – (28), (30), (31) give

$$0 < H < J_1, \qquad m_{n,N} = 0 \qquad (37)$$

$$J_1 < H < 2J_1, \quad m_{n,N} = \left[1 - Q(-H)^{N/2}\right]\mu_{n,N}, \quad \mu_{n,N} = N^{-1}\left[1 + (-1)^n(n' - n)\right] \qquad (38)$$

$2J_1 < H$,

$$m_{n,N} = m_\infty + \int W_\infty \left\{(\rho^n + \rho^{n'})\left[\tilde{C}_\infty(1+\rho) - \rho\right] + \rho^N(1 - 2\tilde{C}_\infty)\right\}dF$$
$$+ (-1)^n \int W_\infty (\rho^n - \rho^{n'})\left[\tilde{C}_\infty(1-\rho) + \rho\right]dF + \mu_{n,N}\left(\int W_\infty \rho^N dF - Q(-H)^{N/2}\right) \qquad (39)$$

Zero magnetization at $0 < H < J_1$ is the consequence of ground-state degeneracy of even chains – they have two ground states in such fields, $S_n = (-1)^n$ and $S_n = (-1)^{n+1}$, so $m_{n,N} = 0$ results from the averaging over them. Applying additionally a small local field $\delta H > 0$ to one of the spins, say, $S_k$, we can lift this degeneracy thus recovering the straight AF order with the unique ground state having $S_k > 0$. This also means that this phase has infinite correlation length because a small local field changes the average magnetization throughout the whole sample. It shows up in the correlation function

$$G_r = \langle S_n S_{n+r}\rangle_0 - \langle S_n\rangle_0 \langle S_{n+r}\rangle_0 = (-1)^r$$

Here $\langle ...\rangle_0$ designates the average over two ground states uninfluenced by disorder. The amplitude of $G_r$ does not fall at large $r$ indicating the infinite correlation length. While $G_r$ describes the reaction of a system on the infinitesimal local perturbations it cannot quantitatively describe the effect of strong ones such as local spin upturn (see below). Yet we can naturally expect that the variations of magnetization caused by a strong local perturbation would also spread throughout all system.

The specific phase appearing at $J_1 < H < 2J_1$ with linearly modulated AF order, see Fig. 4, is also the consequence of ground state degeneracy. It also exists in the ordinary (nonrandom) AF chain with exchange $-J$ and even number of spins when $J < H < 2J$. The mechanism of its appearance is quite simple. Even chains always have in normal AF states one of the end spins, say, $S_0$, pointed opposite to the field. At $H > J$ it would upturn to point along the field thus diminishing the energy by $2(H - J)$. But the simultaneous upturn of three spins at this end, $S_0, S_1, S_2$, gives the same energy gain and generally the same effect results from the upturn of any odd number of spins $S_0, S_1, \ldots, S_{2k}$. Thus we have $N/2$ ground states each having a "kink" - one pair of neighboring spins pointed along the field, see Fig. 5.

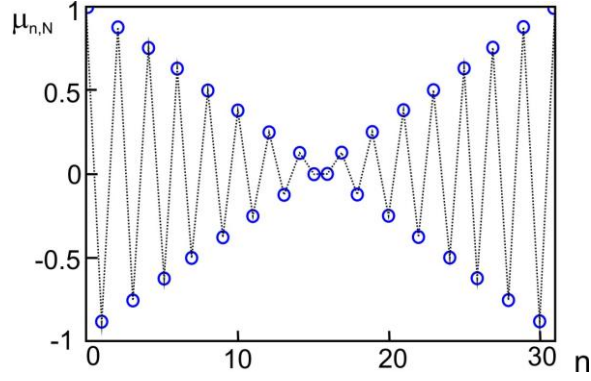

Fig.4. Linearly modulated AF profile $\mu_{n,N}$ of a chain with $N = 32$. Dotted line is a guide to an eye.

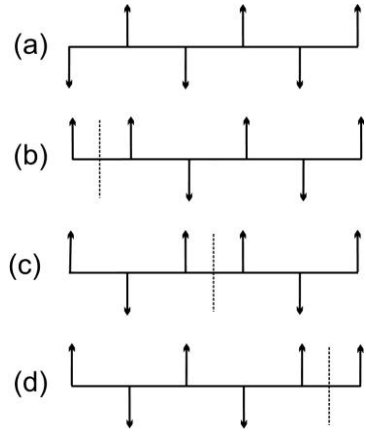

Fig. 5. The kink states (b)-(d) originated from AF order (a) via upturn of the spins to the left from the dashed lines.

The averaging over them gives the "bow-tie" profile shown in Fig. 4. Indeed,

$$m_{n,N} = \frac{2}{N}\sum_{k=1}^{N/2}\left[(-1)^n \vartheta\left(2k - n - \frac{3}{2}\right) + (-1)^{n+1}\vartheta\left(n - 2k + \frac{3}{2}\right)\right]$$

$$= (-1)^n \frac{2}{N}\left\{\frac{N}{2} - 2\left[\frac{n+1}{2}\right]\right\} = \mu_{n,N}$$

In the last expression $\left[(n+1)/2\right]$ is the integer part of (n+1)/2.

This ordering can also be viewed as the boundary effect caused by the end-spin upturn and spreading through the whole sample due to infinite correlation length of the degenerate AF phase in which it originated. So we can expect that such inhomogeneous phases with ordering dependent on the form and size of a sample would also exist in many other systems with infinite correlation length. Among them there are Heisenberg magnets having in the ordered phase infinite transverse correlation length and a number of frustrated magnets in which ground-state degeneracy will also result in the divergence of correlation length at $T = 0$.

It seems that the studies of statistical mechanics of one-dimensional Ising model overlooked somehow the existence of this "bow-tie" phase and for the first time the "kink" states in Fig. 5 was found in the framework of macroscopic Mill's model for finite layered AF [18], see also [19]. This model becomes the AF Ising chain in the limit of infinite anisotropy but, being macroscopic, it does not require the averaging over all kink states. So the authors of Refs. [18, 19] just noted that system with even number of layers can exist in one such state chosen from the set of $N/2$ degenerate ones. Yet in statistical mechanics dealing with statistical ensembles the averaging over degenerate states is inherent procedure. So in its framework taking the limit $T \to 0$ in the standard expressions for AF chain [20] we would obtain $m_{n,N} = \mu_{n,N}$ for even $N$ and $J < H < 2J$. Unfortunately, this is rather hard task which needs to calculate the limits of cumbersome expressions, cf. Ref. [20]. This, probably, explains why this has not been done before.

Our expression for $m_{n,N}$ at $J_1 < H < 2J_1$ (38) differs from that in nonrandom case by the factor $1 - Q(-H)^{N/2}$ only. $Q(-H)$ is the probability to find a "strong" AF bond with $|J| > H$ and there are $N/2$ bonds' positions around which a kink can appear if the bond is a "weak" one, with $|J| < H$, cf. Fig. 5. Hence $Q(-H)^{N/2}$ is the probability that all these positions are occupied by strong bonds and pure AF states are preferable while $1 - Q(-H)^{N/2}$ is the probability that there is at least one weak bond in allowed places and a kink with parallel spins can be created.

One may question the physical observability of the "bow-tie" phase as it needs the ensemble of chains with equal lengths. As physical realization of random AF chains' ensembles the quasi-1d AF and magnetic polymer solutions with vacancies and impurities can be mentioned but they would have a large diversity of chain lengths. Yet this diversity cannot hinder the observation of "bow-tie" phase with neutron diffraction experiments if we have a number of parallel chains having different lengths. The reason for this is that the form of neutron scattering intensity $I(k)$ does not changequalitatively with the chain size. Indeed, $I(k) \sim |m_{k,N}|^2$, $m_{k,N}$ being a Fourier transform of $m_{n,N}$ with discrete transferred wave-vector

$$k = \frac{2\pi l}{N}, \quad l = 0, 1, ..., N-1.$$

In the linearly modulated AF phase under question we have for arbitrary $N$

$$I(k) \sim |\mu_{k,N}|^2 = \left| \frac{2(1-\delta_{k,\pi})}{e^{ik}+1} + \delta_{k,0} \right|^2 = \frac{(1-\delta_{k,\pi})}{\cos^2(k/2)} + 3\delta_{k,0} \qquad (40)$$

Thus $I(k)$ has the same profile for all N, the only difference is in the set of transferred wave-vectors which do not interfere but rather supplement each other as Fig. 6 shows. This makes the observation of the signs of the linearly modulated AF phase feasible in the low-temperature neutron diffraction experiments. In the limit $n \sim n' \to \infty, N \to \infty$ we have $\mu_{n,N} \to 0$. This means that every spin within arbitrary large but finite distance from the center of the chain has average magnetization which tends to zero. This does not mean that there is no phase transition in the thermodynamic limit at $H = J_1$ but only that $m_{n,N}$ is not a correct order parameter for it. As Eq. (40) shows $m_{k,N}$ is a true (multicomponent) order

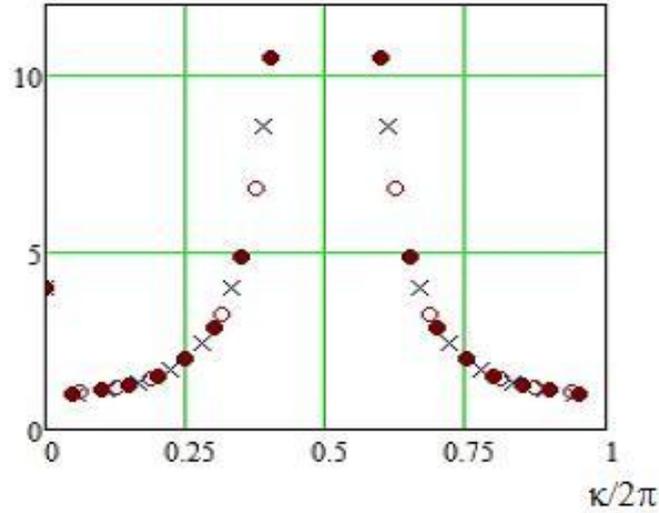

Fig. 6. $I(k)$ for N = 16 (o), N = 18 (x) and N = 20 (●).

parameter and there is a first-order transition between AF and "bow-tie" phases.

The transition at $H = H_c = 2J_1$ in the thermodynamic limit ($n, n', N \to \infty$) has the features similar to those of odd chains in this field – in F phase it is continuous of the second or higher order while in the "bow-tie" phase $m_{k,N}$ sharply drops to zero at the transition point.

In finite samples both transitions become smeared; the factor $1 - Q(-H)^{N/2}$ in (37) rapidly grows from zero to almost unity when H becomes greater than $J_1$, while at $H_c < H$ F order appears gradually in the middle of a chain. Again when $P(J) \to 0$ at $J \to -J_1$ we have near $H_c$

$$m_{n,N} = m_\infty \left(1 + e^{-\kappa N}\right) + (-1)^n \left(1 - m_\infty\right)\left(e^{-\kappa n} - e^{-\kappa n'}\right) + \mu_{n,N}\left(e^{-\kappa N} - Q(-H)^{N/2}\right)$$

with $\kappa$ from Eq.(36). Thus at finite $\kappa$ we have the distribution of average magnetizations similar to that shown in Fig. 3 (b, c) while close to $H_c$ when $\kappa \to 0$ it tends to that of Fig. 4 instead of the straight AF order as in odd chains. When $P(J) \to \infty$ at $J \to -J_1$ magnetization profile shows similar behavior.

We can estimate the temperatures at which present ground state results still hold approximately. The low-T contribution to partition sum is of the order $\exp\left(-\frac{\Delta E_{min}}{T}\right)$ where $\Delta E_{min}$ means the lowest excitation energy above the ground state one. So we may ascertain the validity of present theory at $T \ll \Delta E_{min}$.

For N odd in AF phase ($H < H_c$) the low-energy excitations are the flips of spins directed opposite to the field. Each such flip results in energy change $-2H - 2J_k - 2J_{k-1}$ so

$$\Delta E_{min} = \min_k \left(-2H - 2J_k - 2J_{k-1}\right) > 2\left(H_c - H\right)$$

In F phase ($H > H_c$) the spin flips to the direction opposite to the field results in

$$\Delta E_{min} = \min_k \left(2H + 2J_k + 2J_{k-1}\right) > 2\left(H - H_c\right)$$

Hence in the chains with N odd the present results certainly hold at

$$T \ll |H - H_c|$$

In AF phase ($H < H'_c \equiv J_1$) of even chains the low-energy excitations are the flips of spins considered for the construction of "kink" states, cf. Fig. 5. For them we have

$$\Delta E_{min} = \min_k \left(-2H - 2J_k\right) > 2\left(H'_c - H\right)$$

In bow-tie phase ($H'_c < H < H_c$) along with these kink excitations the ordinary spin flips to the field direction may have the lowest energy so here

$$\Delta E_{min} = \min_k \left(-2H - 2J_k - 2J_{k-1}, 2H + 2J_k\right) > 2\min\left(H_c - H, H - H'_c\right)$$

In F phase we also have single spin flips at low T so in the chains with N even the range of validity of ground state results is

$$T \ll \min\left(|H - H_c|, |H - H_c'|\right)$$

Thus for all chains present results can hold also at sufficiently low T except the vicinity of the transition points.

## 5. Discussion and conclusions.

There are a variety of features specific to the model considered here ($T = 0$, $d = 1$, variation of the external parameter $H$, conjugate to the order parameter $m$ of one of the phases) which distinguish it from a number of conventional smeared first-order transitions. It is still needed to decide to what extent present results are universal. Nevertheless it is the useful example of a strong influence of disorder on a first - order transition in which it becomes (from the F-phase side) a second- or higher-order one with anomalies depending on the bond distribution function.

Thus we have the first definite evidence that critical indices in the emergent second-order transition can be nonuniversal and that the higher-order transitions can appear in the phase coexistence region. Along with this the model exhibits the unexplored possibility that simultaneously the first-order jumps can be preserved on the other side of the smeared transition.

The model also gives the unique opportunity to elucidate the ordering in finite samples which is quite necessary for the description of systems with infinite correlation length. Here such systems are exemplified by the even-site AF chains. The existence of "bow-tie" phase in these chains (either with or without disorder) shows that inhomogeneous size-dependent order can emerge in a system with infinite correlation length due to the influence of boundary effects on the whole bulk ordering. This conclusion is important for the systems with broken continuous symmetry and other degenerate systems such as frustrated magnets where similar phenomena can occur. The evidences of the boundary effects spreading throughout large mesoscopic samples are found in numerical studies of 3d uniaxial AF [21] and 2d Heisenberg AF [22].

Physical realization of considered here continuous distributions with predefined behavior at the upper end may be possibly achieved subjecting AF chains to (artificial) random mechanical stresses which would result in random AF exchanges due to magnetoelastic couplings. Yet to fully conform to the present model these random exchanges between nearest neighbors should be independent. This may be hard to fulfill owing to the long-range nature of deformations caused by random stresses and now it is not clear if the bond correlations can be neglected for some random-bond patterns produced via such mechanism. In any case the present model can be considered as a proper starting point to study more realistic models with bond correlations.

At last we should note that the method presented here may have generalizations for the random-bond Heisenberg, transverse Ising or quasi-1d AF models.

**Acknowledgements**. Author gratefully acknowledges useful discussions with M.P. Ivliev and V.P. Sakhnenko.